\documentclass[12pt, preprint]{aastex} 
\def\beq{\begin{equation}}
\def\eeq{\end{equation}}
\def\simge{\mathrel{%
   \rlap{\raise 0.511ex \hbox{$>$}}{\lower 0.511ex \hbox{$\sim$}}}}
\def\simle{\mathrel{
   \rlap{\raise 0.511ex \hbox{$<$}}{\lower 0.511ex \hbox{$\sim$}}}}
\def\lta{\mathrel{\spose{\lower 3pt\hbox{$\mathchar"218$}}
     \raise 2.0pt\hbox{$\mathchar"13C$}}}
\def\gta{\mathrel{\spose{\lower 3pt\hbox{$\mathchar"218$}}
     \raise 2.0pt\hbox{$\mathchar"13E$}}}
\def\gs{\mathrel{\lower0.6ex\hbox{$\buildrel {\textstyle >}
 \over {\scriptstyle \sim}$}}}
\def\ls{\mathrel{\lower0.6ex\hbox{$\buildrel {\textstyle <}
 \over {\scriptstyle \sim}$}}}

\eqsecnum

\begin{document}
\title{The Large-Scale Bias of the Hard X-ray Background}
\author{S.P. Boughn}
\affil{Department of Physics, Princeton University, Princeton, NJ 08544 and}
\affil{Department of Astronomy, Haverford College, Haverford, PA  19041
sboughn@haverford.edu}
\author{R.G. Crittenden}
\affil{Institute of Cosmology and Gravitation, University of Portsmouth, Portsmouth
PO1 2EG  UK robert.crittenden@port.ac.uk}

\begin{abstract}

Recent deep X-ray surveys combined with spectroscopic identification of the sources
have allowed the determination of the rest-frame $2-8~keV$ luminosity as a function
of redshift.  In addition, an analysis of the $HEAO1~A2~2-10~keV$ full-sky
map of the X-ray background (XRB) reveals clustering on the scale of several degrees.
Combining these two results in the context of the currently favored $\Lambda CDM$
cosmological model implies an average X-ray bias factor, $b_x$, of
$b_x^2 = 1.12 \pm 0.33$, i.e., $b_x = 1.06 \pm 0.16$. These error estimates include
only statistical error; the systematic error sources, while comparable,
appear to be sub-dominant.
This result is in contrast to the large biases of some previous
estimates and is more in line with current estimates of the optical bias of $L_*$ galaxies.

\end{abstract}


\section{Introduction}

An important test of any cosmological model is that it be consistent with 
the observed distribution of matter in the universe.  Since our primary knowledge
of this distribution comes from observations of galaxies, it is 
essential to understand the extent to which galaxies trace the matter density.  
This relationship is usually quantified by a bias factor which relates fluctuations in 
the galaxies to those in the dark matter.  It
is complicated by the fact that the relation between the luminosity of a galaxy 
or groups of galaxies and the underlying distribution of matter can depend on the 
type of galaxy, the spectral band of the observation, the redshift $z$, and the scale
length on which the comparison is made.  However, such complications are also 
opportunities in that models of galaxy formation must successfully reproduce these
differences. 

The standard definition of the bias factor, $b$, is the ratio of the fractional galaxy 
density fluctuations to the fractional matter density fluctuations, i.e.,
\begin{equation}
b = {\delta \rho_g / \rho_g \over \delta \rho / \rho}
\end{equation}
where $\rho_g$ is the mean density of galaxies, $\rho$ is the mean density 
of matter and $\delta$ indicates the $rms$ fluctuations of the densities about 
these means.  
If galaxies formed early ($z >  1$), as appears to be the case 
(e.g., Ellis 1997), then there are
good reasons to expect that, for linear density perturbations
(i.e., $\delta \rho / \rho \ll 1$) on large scales in the nearby ($z \ls 1$)
universe, galaxies should be relatively unbiased ($b \rightarrow 1$) tracers of the density field
(Fry 1996, Tegmark \& Peebles 1998); however, this assertion must be tested.

Here we focus on determining the bias of the hard X-ray background (XRB),
which is
known to be dominated by distant ($z \ls 2$) active extragalactic galaxies and 
so provides a probe of the bias on large scales.   
The observed clustering of the XRB, when combined with what is 
known about the level of perturbations and the cosmological model from CMB 
observations, allows us to place relatively strong constraints on the
X-ray bias.

Previous determinations of X-ray bias have resulted in a wide range of values,
$1 < b_x < 7$ (see Barcons et al. 2000 and references therein).  Some 
spread in the estimates is to be expected; e.g., at lower energies 
X-ray emission is dominated by clusters of galaxies, and so are expected to be as 
highly biased as clusters themselves (Bahcall \& Soneira 1983). 
However, another major contribution to the uncertainty in the bias estimates is 
the lack of accurate determinations of the clustering of various X-ray sources.  
Two of the lower estimates of 
X-ray bias are from Treyer et al. (1998) who found that $b_x \sim 1$ to $2$ from a 
low order multipole analysis of the HEAO1 A2 data set and Carrera et al. (1998) 
who found that the ratio of the X-ray bias of AGN to that of IRAS galaxies to be 
$0.8 \ls b_x/b_I \ls 1.7$ from ROSAT observations.  The remaining uncertainty
in these determinations arose from uncertainties in both the
X-ray luminosity function (LF) and in the cosmological model.  Knowledge of both of these
has improved dramatically in the last year and this is largely 
responsible for the improved accuracy of the estimate of $b_x$ in this paper.

\section{Clustering in the HEAO1 A2 $2-10~keV$ X-ray Map}

We recently presented evidence of large-scale clustering in the HEAO1 A2 $2-10~keV$ 
full-sky map of the hard XRB on angular scales of $\ls 10^{\circ}$
(Boughn, Crittenden, \& Koehrsen 2002).  
Before computing the correlations, local sources of the X-ray background were 
removed from the map.  
The map was masked so as to eliminate strong, nearby X-ray sources with fluxes 
exceeding $3 \times 10^{-11}erg~s^{-1}cm^{-2}$.  In addition, all regions within $20^{\circ}$ 
of the Galactic plane or within  $30^{\circ}$ of the Galactic center were masked.  The
map was also corrected for a linear time drift of the detectors, high Galactic latitude
diffuse emission, emission from the local supercluster, and the Compton-Getting dipole.
The latter components were fit to and then removed from the map.  
Without these cuts and corrections, the correlations are dominated by a 
few strong point sources and large-scale diffuse structures in the map.

\clearpage

\begin{figure}
\centerline{
\plotone{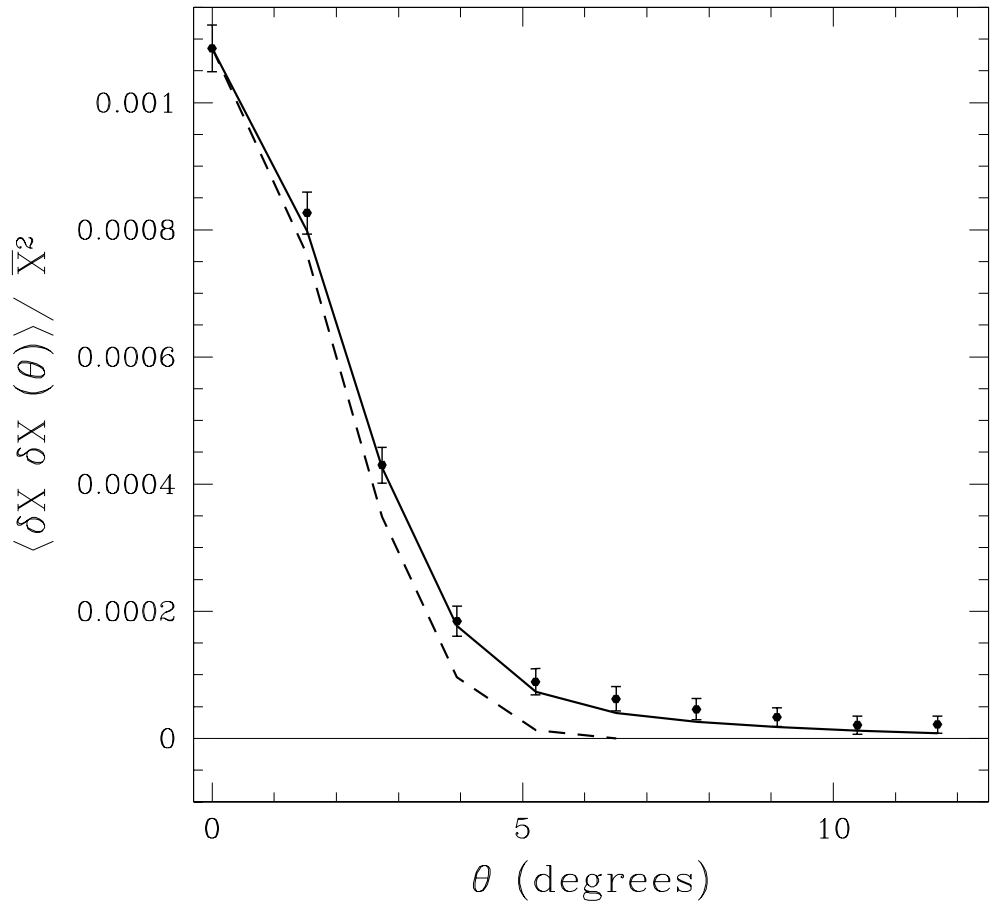}}
\caption{The auto-correlation function of the HEAO1 A2 map
with bright sources and the Galactic plane removed. The dashed 
curve is the form expected from beam smearing due to the PSF
of the map while the solid curve includes a contribution due to
clustering in the XRB.} 
\label{fig:acf}
\end{figure}

\clearpage
 
Figure 1 is a plot of the intensity angular 
correlation function (ACF) given by
\begin{equation}
ACF(\theta) = {1 \over {\bar{I}^2 N_{\theta}}} \sum_{i,j} (I_i -\bar{I})
(I_j -\bar{I})
\end{equation}
where the sum is over all pairs of map pixels, $i,j$, separated by an angle 
$\theta$, $I_i$ is the intensity of the $ith$ pixel,
$\bar{I}$ is the mean intensity, and $N_{\theta}$ is the number of 
pairs of pixels separated by $\theta$.  Photon shot noise only appears
in the $\theta = 0^\circ$ bin and has been removed. The highly correlated error bars 
were determined from 1000 Monte Carlo trials in which the pixel intensity distribution was 
assumed to be Gaussian with the same ACF as in the figure.  The dashed curve represents 
the expected functional form of the contribution to the ACF due to telescope beam smearing 
of a random distribution of uncorrelated sources normalized to the ACF(0) point.  It
represents the profile that is expected if there were no intrinsic correlations in
the XRB.  The point spread function (PSF) of the map is due to pixelization 
and to the finite telescope beam and was accurately determined from the profiles of 
60 bright, nearby point sources.  
It is clear from Figure 1 that the XRB possesses intrinsic (i.e., not due to beam smearing) 
correlated structure out to angular scales of $\sim 10^{\circ}$. 
Full details of the analysis are discussed in Boughn, Crittenden, \& Koehrsen (2002).

\section{The $2-10~keV$ X-ray Luminosity Function}

In order to determine the X-ray bias factor $b_x$ from the measured ACF, it is essential
to know from which redshifts the X-ray fluctuations originate; the underlying 
density fluctuations grow quickly, so it is important that they be compared to the 
X-ray fluctuations at the same redshifts.   
This requires understanding the contribution to the $2-10~keV$ X-ray LF 
as a function of redshift.   
However, the HEAO1 A2 observations are total intensity 
measurements of the hard XRB with no information as to 
the fluxes or redshifts of individual sources, so we must infer the LF by other means.   
Recently the 
$Chandra$ satellite has made possible large, faint hard X-ray
surveys with measured redshifts.  Cowie et al. (2003) and Steffen et al. (2003) have 
combined $Chandra$ sources with brighter sources from $ASCA$ (Akiyama et al. 
2000) and $ROSAT$ (Lehmann et al. 2001) to determine the redshift evolution of the 
$2-8~keV$ LF with few assumptions about the character of the sources.
The incompleteness uncertainty in the redshift dependence of the volume X-ray emissivity 
is estimated to be a factor $\ls 2$ at any redshift.
The spectroscopically identified sources comprise $75\%$ 
of the total $2-8~keV$ X-ray intensity; the dominant uncertainties result from 
the unknown redshift distribution of the unidentified sources.  

\clearpage

\begin{figure}
\centerline{
\plotone{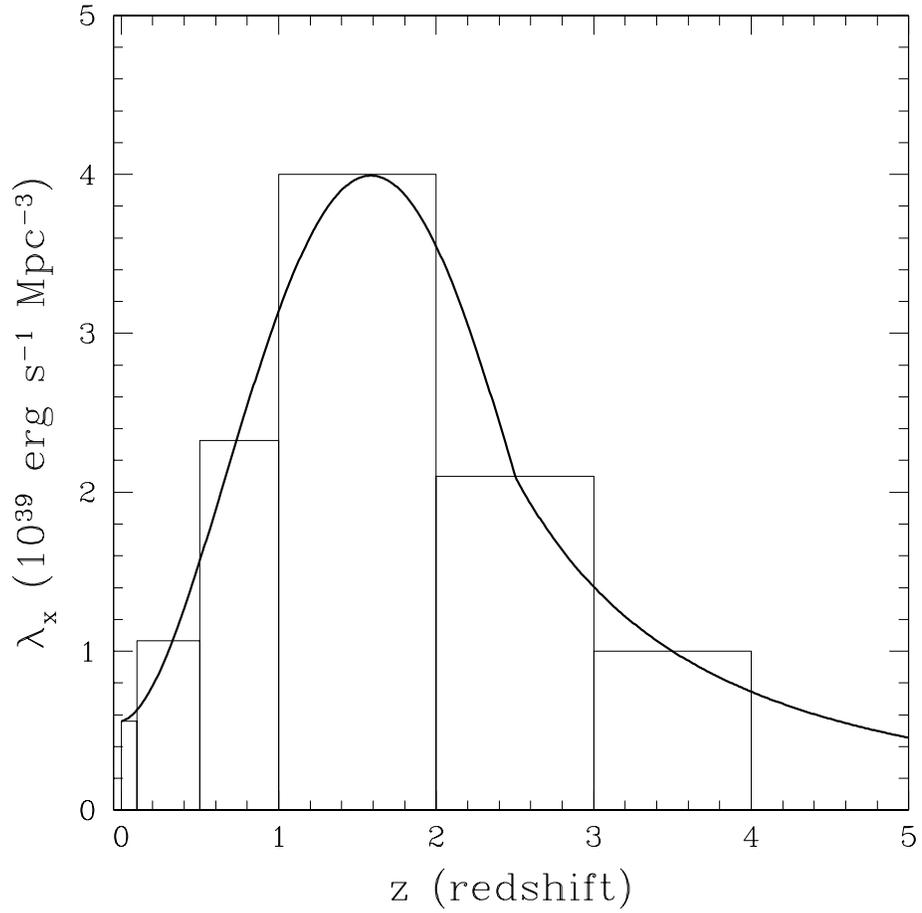}}
\caption{The volume emissivity as a function of redshift.  The local value ($z=0$) is the measurement 
of Miyaji et al. (1994).  The low redshift points are derived from the data of Steffen et al. (2003)
while the high redshift ($z>1$) come from Cowie et al. (2003). 
}
\label{fig:emiss}
\end{figure}

\clearpage

Emissivity as a function of redshift, $\lambda_x(z)$, 
is plotted in Figure 2.  
The $2-8~keV$ emissivity in the redshift range $1 < z < 4$ is taken to be that estimated 
by Cowie et al. (2003) using ROSAT data.  In the range $0.1 < z < 1.0$, we use the 
emissivity implied from the luminosity function of Steffen et al. (2003).
Finally, for $z = 0$, we use the value of the local emissivity from Miyaji et al. (1994).
The models discussed below are based on a polynomial fit which passes through the data points; 
however, the results are largely independent of the 
details of the fitted function.

\clearpage

\begin{table*}[ht]
\begin{center}
\begin{tabular}{c||c|cccc}
\multicolumn{1}{l}{} & \multicolumn{1}{c}{Observed} & \multicolumn{1}{l}{Best}
& \multicolumn{1}{l}{Low $z$} & \multicolumn{1}{l}{High $z$}
& \multicolumn{1}{l}{Ueda {\it et al.}}\\
\cline{1-6}
& & & & & \\
f($z<1$)   &   54\%/58\%  & 57 \%  &  67 \% & 48 \% & 52 \% \\
$b_x$   &  -  & 1.06  &  0.86 & 1.36 & 1.12 \\
                                                                                                         
\end{tabular}
\caption{Properties of four models of X-ray emissivity: fraction of the intensity
arising from $z < 1$, $f(z < 1)$) and implied bias, $b_x$.  See text for details.
}
\end{center}
\label{tab:dipole}
\end{table*}

\clearpage

The HEAO data is band limited and the X-rays detected at high redshifts have  
larger rest frame energies than they do locally.  Therefore, 
in order to apply K-corrections to the observed intensities and to transform 
from $2-8~keV$ emissivities to the $2-10~keV$ values appropriate for the HEAO map,
we must make an assumption about the frequency and redshift dependence of the XRB.
If the ISM column density in front of an AGN is large enough
($N_H \gs 10^{21} cm^2$) the observed spectrum will be hardened by
photo-electric absortion.  At high redshifts, the rest frame 
energies of the detected X-rays are relatively large and the effect of 
photo-electric absorption is less.  Therefore, for a given column density, 
sources at high redshift will appear softer than their low redshift
counterparts.
As a crude approximation of this affect
we assumed a photon spectral index $\Gamma(z) = 1.2+0.2z$ where $dN/dE \propto E^{-\Gamma}$
is the number spectrum of the photons of energy $E$.  This roughly describes
the redshift dependence of sources with an intrinsic spectral index of $\Gamma = 1.8$
subject to photoelectric absorption by column densities of $N_H \sim 10^{22} - 10^{23}~cm^{-2}$.
Furthermore, the flux weighted average spectral index is $\Gamma = 1.4$ as is observed.
The difference between this model and one that assumes a constant
spectral index of 
$\Gamma(z) = 1.4$ is not large in the sense that the intensity
distributions, $dI/dz$, of the two models fall well within the range of the extreme models of
Figure 3. 
It is also possible that higher redshift AGN are more heavily absorbed (e.g. Worsley et al. 2004); 
however, such effects are not included in some models of the XRB (e.g., Ueda et al. 2003).
If the absorption levels were higher at higher redshifts, this would reduce the evolution 
of the spectral index, i.e., lessen it's dependence on redshift.   The fact that we
observe more energetic rest frame photons is balanced by the fact that
the sources are more absorbed.  While neither of the
models (constant $\Gamma(z) = 1.4$ or $\Gamma(z) = 1.2+0.2z$) is likely 
to accurately describe the actual spectrum, the fact that the biases 
of these models are within a few percent of each other indicates that uncertainty in 
the redshift dependence of the spectral index of the XRB is not an important source
of systematic error.

\clearpage

\begin{figure}
\centerline{
\plotone{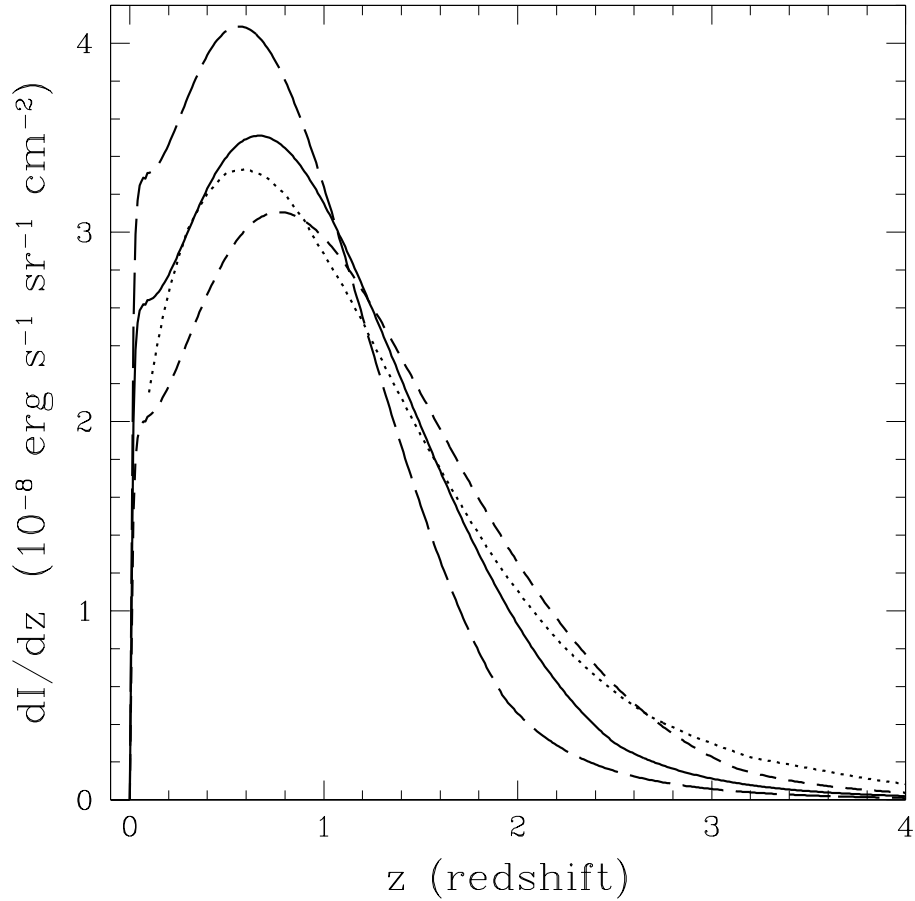}}
\caption{ Four different models for the contribution to the 
x-ray luminosity as a function of redshift.  The solid line is our best estimate given the 
volume emissivity.  The long (short) dashed model results from pushing the emissivity 
distribution to lower (higher) redshifts.  Finally, the dotted line represents the model of 
Ueda et al. (2003), which is not constrained below $z = 0.1$.  
}
\label{fig:dfdz}
\end{figure}

\clearpage

It is straightforward to compute the intensity distribution, $dI/dz$, 
from $\lambda_x(z)$ in the context of the  $\Lambda CDM$ cosmological model used 
by Cowie et al (2003) ($\Omega_m = 1/3$, $\Omega_{\Lambda} = 2/3$, and 
$H_0 = 65~km~s^{-1}Mpc^{-1}$).  While this model is somewhat different from that
currently favored by the $WMAP$ satellite data (Spergel et al. 2003), $dI/dz$ is a 
directly observable quantity that is independent of the cosmological model.
The $dI/dz$ resulting from our canonical emissivity model is given by
the middle solid curve in Figure 3 where the normalization is fixed by the LF of Cowie 
et al. (2003) and Steffen et al. (2003).  

This profile implies that the bulk of the XRB 
arises at much lower redshifts than previously thought (e.g., Comastri et al. 1996) as was 
first pointed out by Barger et al. (2001).  For this particular model, $57\%$ of the 
$2-10~keV$ background
arises from redshifts less than 1.  This is in agreement with the recent
observations of Barger et al. (2003) that indicate $54\%$ of the spectroscopically
identified $2-8~keV$ intensity arises at $z < 1$.  This increases to $58\%$ when photometric 
redshifts are included (Barger et al. 2002; Barger et al. 2003).
The total integrated intensity of our canonical model,
$5.6 \times 10^{-8}erg~s^{-1}cm^{-2}sr^{-1}$, lies between and is consistent with both the $HEAO$ 
estimate of $5.3 \times 10^{-8}erg~s^{-1}cm^{-2}sr^{-1}$ (Marshall et al. 1980; Gruber et al. 1999) 
and that estimated from $ASCA$ satellite data (Gendreau et al. 1995; Kushino et al. 2002), 
$6.4 \times 10^{-8}erg~s^{-1}cm^{-2}sr^{-1}$.  In any case, the current
analysis only requires the functional form of $dI/dz$ and not the overall normalization.
Finally, the intensity weighted spectral index of the model, $\bar{\Gamma} = 1.40$, is the same as that 
observed for the hard XRB (Marshall et al. 1980; Gendreau et al. 1995).

In order to test the sensitivity of the implied X-ray bias to the LF, 
we consider three alternative models of $dI/dz$.  The upper dashed curve 
in Figure 3 has been weighted to low $z$ by squeezing (in redshift) the canonical 
emissivity by a factor of $0.8$ while fixing the local emissivity to be the 
$1~\sigma$ upper limit of Miyaji et al. (1994).
This model is fairly extreme, as it  overestimates the intensity coming from low redshifts.
(See Table 1 for a summary of the properties of the various X-ray models).
The lower dashed curve in Figure 3 was weighted to high $z$ by stretching the
canonical emissivity by a factor of $1.3$ while fixing the local emissivity to be the 
$1 \sigma$ lower limit of Miyaji et al. (1994).  This model significantly underestimates the 
intensity coming from $z<1$. 
Finally, the dotted curve in Figure 3 is from the recent AGN synthesis model of
Ueda et al. (2003).  Unfortunately, their model 
of $dI/dz$ did not extend below $z = 0.1$ and 
our results below depend somewhat on the behaviour assumed for low redshifts.

\section{Matter Fluctuations in a $\Lambda CDM$ Universe and X-ray Bias}

Given the X-ray luminosity function, 
the linear bias factor can be inferred from the cosmological
model, but only if the time dependence and scale dependence of the bias
are known.  
In our analysis we will assume both redshift and scale
independence of the X-ray bias.
Given our nominal intensity distribution, 
and assuming the standard $\Lambda CDM$
cosmology, the dominant contribution to the ACF on angular scales of a few 
degrees comes from structures with redshifts, $0.03 < z < 0.5$, which
correspond to linear scales of from approximately $10~Mpc$ to $200~Mpc$.  
This is a strong indication that we are observing clustering in the linear 
regime and so can use the straightforward analysis of the growth of linear 
structures in a $\Lambda CDM$ universe. 


Using the current $WMAP$ $\Lambda CDM$ parameters (Spergel et al. 2003), 
i.e., $\Omega_m = 0.27$, $\Omega_{\Lambda} = 0.73$, and 
$H_0 = 71~km~s^{-1}Mpc^{-1}$, it is straightforward to compute a projected 
matter ACF with the same redshift distribution as for the canonical model 
(e.g., Boughn, Crittenden, \& Turok 1998).  If our assumptions about 
the bias are correct, the intrinsic X-ray ACF should have the same shape as 
the normalized matter ACF, with a relative amplitude given by the  
square of $b_x$, the X-ray bias factor.  

The observed ACF also contains 
components due to beam smearing of uncorrelated X-ray sources and photon 
shot noise, the latter of which is uncorrelated and, therefore, only 
contributes to the ACF at $\theta = 0$. Therefore, any fit to the full data set 
must include these three components.  At  $\theta = 0^{\circ}$ the ACF
is dominated by beam smearing and photon shot noise while above 
$\theta = 12^{\circ}$ the signal to noise is small.  The solid curve
in Figure 1 is the two parameter, maximum likelihood fit to the data in 
the range $2.5^{\circ} < \theta < 12^{\circ}$.  The implied X-ray bias is 
${b_x}^2 = 1.12 \pm 0.33$ ($1~\sigma$ error) or $b_x = 1.06 \pm 0.16$ 
with a $\chi^2$ of 4.6 for 6 degrees of freedom.  Since the distribution of 
errors in the ACF is to a good approximation Gaussian, the statistical error
attached to ${b_x}^2$ as well as the $\chi^2$ of the fit have the usual
interpretations.  The error indicated for $b_x$ represents the $68\%$
confidence interval; however, this error is not Gaussian. 
The signal to noise of the data point at $5.2^{\circ}$
is $4~\sigma$ and a variety of fits (see below) of ${b_x}^2$ to the ACF 
indicate statistical significances between $3$ and $4~\sigma$.  

We performed a variety of other fits to the data to check the robustness of
our estimate of ${b_x}^2$.  A three parameter fit to the data in the full interval 
($0^{\circ} < \theta < 12^{\circ}$) gives $b_x = 0.96 \pm 0.16$. A one parameter
fit for the large angle correlations ($5.2^{\circ} < \theta < 12^{\circ}$), where  
the beam smearing component is nearly negligible, gives $b_x = 1.20 \pm 0.14$;   
even a fit to the single datum at $5.2^{\circ}$ yields a
consistent value of $b_x = 1.25 \pm 0.16$, though it is probably mildly
contaminated by the beam smearing component.  Following our previous 
work (Boughn, Crittenden, \& Koehrsen 2002), we also modeled the clustering term
as a power law, $\propto 1/\theta^{\alpha}$, with $0.8 < \alpha < 1.6$.  These fits
varied in amplitude; however, at $\theta = 4.5^{\circ}$ all of the fits agreed 
to within a few percent.  Normalizing the model clustering ACF to this level 
implies a bias of $b_x = 1.06 \pm 0.17$, also consistent with our canonical
fit.  The reduced $\chi_{\nu}^2$'s for these fits are all $\sim 1$ and
the fits are all consistent with each other.

The process of fitting for large-scale, diffuse components and then removing them
from the $HEAO$ map, results in some attenuation of the ACF on angular scales 
$\gs 10^{\circ}$.  These factors were determined from the same Monte Carlo
trials that were used to determine the statistical errors and the fits were adjusted
accordingly.  Even if these factors are ignored, the fit value of $b_x$ changes by 
only $3\%$.  

To evaluate the level of uncertainty due a systematic error
in the intensity distribution of the XRB, the two ``extreme''
models of Figure 3 were also fit to the data in the 
$2.5^{\circ} < \theta < 12^{\circ}$ interval.  The biases resulting from these
two fits are $b_x = 0.85 \pm 0.13$ for the low z model and $b_x = 1.36 \pm 0.21$
for the high z model.  Since these models are somewhat exaggerated, we conclude that 
they represent lower and upper limits of systematic errors due to uncertainty
in $dI/dz$.  A fit to the Ueda et al. (2003) model indicated in Figure 3
results in a similar value of the bias, though the precise results depend on how the 
model is extended to low redshifts ($0 < z < 0.1$).  
If this model is extended so that the low z behavior is not
allowed to fall below that implied by Miyaji et al. (1994),  
$dI/dz = 2.7 \times 10^{-8} erg~s^{-1}cm^{-2}sr^{-1}$, then the fit value of $b_x$ becomes 
$1.12 \pm 0.17$, which is consistent with that implied by our canonical model.  
If instead, we use a linear extrapolation to low redshifts, the bias can be 
somewhat ($\sim 15 \%$) higher, but the local emissivity of this model  
would be nearly $2\sigma$ below that implied by Miyaji et al. (1994).

The ACF on large angular scales is quite sensitive to the contribution of low $z$ 
sources (roughly half the ACF at $\theta = 4.5^{\circ}$ is due to sources with 
$z \ls 0.1$), 
so any error in estimating the low redshift cutoff in $dI/dz$ could affect the results 
dramatically.   By masking sources stronger than 
$3 \times 10^{-11}erg~s^{-1}cm^{-2}$ we effectively 
truncate the intensity distribution at low redshifts.  The truncated 
profiles were determined from the flux cut and the local luminosity function 
of Steffen et al. (2003). If the value of the flux cut is in error due to, 
for example, a difference in normalizations of the source catalog used to make the cuts 
(Piccinotti et al. 1982) and the Steffen et al. luminosity function, then this 
would affect the cutoff redshift and  would be 
translated to an error in the predicted ACF. 
In the extreme limit of no flux cut, i.e., no truncation of the $dI/dz$ 
profile, the implied X-ray bias is $b_x = 0.90$.  In the other extreme, i.e.,
a flux cutoff of $1 \times 10^{-11}erg~s^{-1}cm^{-2}$, the implied bias is $b_x = 1.13$.
Therefore, it is unlikely that inaccuracy in characterizing the flux
cut is the source of significant systematic error.

Potentially more problematic is the redshift distribution of the unresolved component of the
XRB.  Worsley et al. (2004) found that above $7~keV$ only $\sim 50\%$ of the XRB is resolved;
although, this conclusion must be tempered somewhat by the fact that the brightest sources 
they considered (in the Lockman Hole) have fluxes of $\sim 10^{-13}~erg~s^{-1}~cm^{-2}$.
Sources brighter than this contribute to the whole-sky XRB and
they conclude that the true resolved fraction may be 10 to 20\% higher.
Even though only $\sim 20\%$ of the counts in the HEAO passband comes from photons with energies
above $7~keV$, an unresolved component can still significantly affect the estimate of the bias.
As a pessimistic case, we ignore the bright source correction and assume a $30\%$ unresolved 
component below $5~keV$.  In this case roughly $1/3$ of the 2-10 $keV$ XRB is unresolved.  If 
this unresolved component is distributed in redshift like the resolved component, then there is 
no change in the implied bias.  On the other hand, if the unresolved component is entirely due
to sources at high redshift where it does not contribute to the ACF signal, then the implied
bias will be $50\%$ higher than our canonical value.  If instead the unresolved 
component is due to sources at low redshift, $z < 1$, then the implied bias will be 
$20\%$ lower than our canonical value.  These fall somewhat outside our two ``extreme'' values 
in Table 1 and so provide a caveat to those estimates of the limits of systematic error.
However, if only half of the unresolved component is located at high (low) redshifts and the 
other half is distributed like the resolved component, then the implied bias is only $20\%$
($11\%$) higher (lower) than our canonical value, well within the limits of Table 1.

It is difficult to quantify all possible systematic errors; however, considering that 
the above ``extremes'' result in errors of the same order as the statistical error in the
fit, we conclude that the total systematic error is no larger than the statistical error
quoted.

\section{Discussion}

We have determined the X-ray bias of the hard XRB  
assuming it is time
(i.e., redshift) and scale independent.  These
assumptions are probably quite reasonable since the mean redshift weighting
of the X-ray ACF is quite low, $z \sim 0.1$, and the linear scales probed 
by the ACF are quite large ($10~Mpc$ to $200~Mpc$).  Even if these assumptions
are violated to some extent, $b_x$ can still
be interpreted as an `average' X-ray bias.  There are
several types of sources that contribute to the XRB, including quasars, Seyfert
galaxies, LINERS, and clusters of galaxies, and the implied value of the bias
must be considered to be an average over all these sources. However,
the dominant contribution to the XRB is most likely to be moderately
active AGN (Cowie et al. 2003), so $b_x$ should be
representative of the bulk of the sources of the XRB.

With these caveats in mind, we find an X-ray bias of 
$b_x^2 = 1.12 \pm 0.33$, i.e., $b_x = 1.06 \pm 0.16$ (statistical error only).
This error includes photon shot noise, fluctuations in the XRB from 
beam smearing, and the clustering of the XRB itself.  The fits of $b_x$ for two extreme 
models of $dI/dz$ indicate that the uncertainty due to our ignorance of the X-ray 
luminosity function is likely less than the statistical error.  Other possible sources of 
systematic error also seem small.  
We conclude that the hard XRB is a largely
unbiased tracer of the matter distribution on
large scales.  This is consistent with current models of large-scale,
late time galaxy biasing (Benson et al. 2000; Tegmark \& Peebles 1998).  
In addition, the latest studies of the
clustering of $\sim L_*$ galaxies on $\sim 100~Mpc$ scales indicates that these
objects are also unbiased tracers of matter.  Verde et al. (2002) found that,
on scales of $\sim 7$ to $\sim 40~Mpc$,
$b = 1.04 \pm 0.11$ for $1.9~L_*$ galaxies in the $2dF$ survey with a mean
redshift of $z = 0.17$.  Using a different analysis of the same data, 
Lahav et al. (2002) found that $b = 1.20 \pm 0.11$ on scales of
$\sim 20$ to $\sim 150~Mpc$. Both of these results are consistent with early findings
from the SDSS and 2MASS surveys that imply linear bias factors on the order of unity
(Tegmark et al. 2002; Miller et al. 2003).  It should not be surprising that 
the XRB and galaxy biases are similar since $L_*$ galaxies 
are closely associated with the moderately active AGN that comprise the bulk of the 
hard XRB (e.g., Barger et al. 2003; Miller et al. 2003).

If these estimates are accurate, then the X-ray bias factor in the linear regime 
is now much better determined.  
The hard XRB background appears to be an excellent tracer of the large-scale 
distribution of matter, making it a useful tool for understanding
the evolution of structure in the universe.    
One example of the importance of determining galaxy biases (and indeed the 
driving motivation for this work) is to aid in the interpretation of recent 
detections of correlations of galaxies with the cosmic microwave
background (CMB).  
We (Boughn \& Crittenden 2004)
have detected a correlation of the $2-10~keV$ XRB with 
$WMAP$ satellite map of the cosmic microwave background (Bennett et al. 2003), 
and there have been correlations observed with a number of other galaxy surveys 
(Nolta et al. 2003; Scranton et al. 2003; Fosalba, Gaztanaga,\&
Castander 2003; Afshordi, Loh, \& Strauss 2003). These correlations have been 
interpreted as the detection of the integrated Sachs-Wolfe ($ISW$) effect
(Sachs \& Wolfe 1967).  If confirmed, they would constitute 
an important test of the $\Lambda CDM$ cosmological model and provide further evidence 
of the existence of a substantial amount of ``dark energy'' in the universe 
(Crittenden \& Turok 1996).  

\begin{acknowledgments}

We would like to acknowledge Keith Jahoda who is responsible for constructing
the HEAO1 A2 X-ray map and who provided us with several data-handling
programs.  We also thank Greg Koehrsen for noise analysis programs.
RC acknowledges financial support from a PPARC AF fellowship.

\end{acknowledgments}


\begin{references}
\reference{} Afshordi,N., Loh. Y.S.,\& Strauss, M. A. 2003, astro-ph/0308260
\reference{} Akiyama, M. et al. 2000, ApJ, 532, 700.
\reference{} Bahcall, N. A., \& Soneira, R. M. 1983, ApJ, 270, 20.
\reference{} Barcons, X., Carrera, F.J., Ceballos, M.T. \& Mateos, S. 2000,
Invited review presented at the Workshop X-ray Astronomy'99: Stellar
Endpoints, AGN and the Diffuse X-ray Background (also astro-ph/0001182).
\reference{} Barger, A. J. et al. 2001, AJ, 122, 2177.
\reference{} Barger, A. J. et al. 2002, AJ, 124, 1839.
\reference{} Barger, A. J. et al. 2003, AJ, 126, 632.
\reference{} Bennett, C. L et al. 2003 ApJS, 148, 1.
\reference{} Benson, A. J., Cole, S., Frenk, C.S., Baugh, C. M., \& Lacey, C. G.
2000, MNRAS 311, 793.
\reference{} Boldt, E. 1987, Phys. Rep., 146, 215
\reference{} Boughn, S. 1999, ApJ, 526, 14
\reference{} Boughn, S. \& Crittenden, R. 2003, Nature, 427, 45.
\reference{} Boughn, S., Crittenden, R., \& Koehrsen G. 2002, ApJ, 580, 672.
\reference{} Boughn, S., Crittenden, R. \& Turok, N. 1998, New Astron., 3,
275 
\reference{} Carrera, F. et al. 1998, MNRAS, 299, 229.
\reference{} Comastri, A., Setti, G., Zamorani, G. \& Hasinger, G.
 1995, A \& A, 296, 1
\reference {} Cowie, L. L., Barger, A. J., Bautz, M. W., Brandt, W. N., Garmire, G. P.
2003, ApJ, 584, L57. 
\reference{} Crittenden, R. \& Turok, N. 1996, PRL 76, 575
\reference{} Ellis, R. S. 1997, Annu. Rev. Astron. Astrophys., 35, 389.
\reference{} Fosalba, P., Gaztanaga, E., \& Castander, F. 2003, astro-ph/0307249
\reference{} Fry, J. N., 1996, ApJ, 461, L65. 
\reference{} Gendreau, K. C. et al. 1995, PASJ, 47, L5
\reference{} Gruber, D. E., Matteson, J. L., Peterson, L. E.,\& Jung, G. V. 1999,
ApJ, 520, 124.
\reference{} Kushino, A. et al. 2002, PASJ, 54, 327.
\reference{} Lahav, O. et al. 2002, MNRAS, 333, 961.
\reference{} Lehmann, I. et al. 2001, A\&A,371, 833.
\reference{} Maller, A. H., McIntosh, D. H., Katz, N., \& Weinberg, M. D. 2003,
ApJ, in press (astro-ph/0304005).
\reference{} Marshall, R. E. et al. 1980, ApJ, 235,4.
\reference{} Miller, C. J., Nichol, R. C., Gomez, P., \& Hopkins, A. 2003,
ApJ, in press (astro-ph/0307124).
\reference{} Miyaji, T., Lahav, O., Jahoda, K., \& Boldt, E. 1994, ApJ 434, 424.
\reference{} Nolta, M.R. et al. 2003, ApJ, in press
 Boldt, E. \& Piran, T. 2000, ApJ, 544, 49
\reference{} Piccinotti, G., Mushotzky, R. F.,  Boldt, E. A.,
 Holt, S. S., Marshall, F. E., Serlemitsos, P. J.  \&
 Shafer, R. A. 1982, ApJ, 253, 485
\reference{} Sachs, R. K. \& Wolfe, A. M. 1967, Ap J, 147, 73.
\reference{} Scranton, R. et al. 2003, astro-ph/0307335
\reference{} Spergel, D. N. et al. 2003, ApJS, 148, 175.
\reference{} Steffen, A.T., Barger,A.J., Cowie, L.L., Mushotzky, R.F., Yang, Y.
2003, ApJ, 596, L23.
\reference{} Tegmark, M. \& Peebles, P. J. E. 1998, ApJ, 500, L79
\reference{} Tegmark, M. et al. 2002, ApJ, 571, 191.
\reference{} Treyer, M. A., Scharf, C. A., Lahav, O., Jahoda, K.,
 Boldt, E. \& Piran, T. 1998, ApJ, 509, 531
\reference{} Ueda, Y., Akiyama, M., Ohta, K., \& Miyaji, T. 2003, ApJ, 598, 886.
\reference{} Verde, L. et al. 2002, MNRAS, 335, 432.
\reference{} Worsley, M. A. et al. 2004, astro-ph/0404273.
\end{references}
\end{document}